# Transport properties of nuclear matter from anomalous fission yields

**K.-H. Schmidt[1*], Ch. Schmitt[2,3], A. Heinz[4]**

[1]Rheinstr. 4, 64390 Erzhausen, Germany.
[2]Institut Pluridisciplinaire Hubert Curien, CNRS/IN2P3-UDS, 67037 Strasbourg Cedex 2, France.
[3]Institute of Nuclear Physics Polish Academy of Sciences, 31-342 Krakow, Poland.
[4]Department of Physics and Astronomy, Chalmers University of Technology, 41296 Gothenburg, Sweden.

**Abstract**

In nuclear fission, a heavy nucleus splits into two fragments, driven by the Coulomb repulsion between the positively charged protons. The fission process is governed by the potential energy and basic transport properties of nuclear matter like inertial mass and viscosity. Both inertia and viscosity induce a delay, the so-called relaxation time, in the response towards statistical equilibrium of the mass-asymmetry degree of freedom on the fission path. We show that the conditions in the mass-asymmetry degree of freedom near the second barrier are preserved for all systems at excitation energies above a certain threshold. Anomalies that were hitherto unexplained appear in the fission yields and total kinetic energies at lower energies for trajectories, which can reach scission only by quantum-mechanical tunneling through the potential beyond the second barrier. This indicates that the relaxation time in the mass-asymmetry degree of freedom for classically allowed trajectories is longer than the dynamical saddle-to-scission time. This finding is the central result of our work that provides novel information on the transport properties of nuclear matter. Possible scenarios that explain this finding are discussed. These are either a long oscillation time due to a large influence of inertia or a strong friction resulting from a large viscosity. The first option is in severe conflict with the widely used assumption that the role of collective inertia in fission dynamics is negligible, while both options contradict the widespread assumption of local statistical equilibrium in all collective degrees of freedom along the fission path.

## Introduction

The properties and the behaviour of the matter in our environment that we normally experience are determined by the electric forces between the electrons of the atoms and molecules on an energy scale of a few electron Volts (eV). On a scale of about 1 million times higher energies, dynamical processes in and between atomic nuclei like fission and fusion, come into play. They are governed by the nuclear forces, which are complex and not fully understood. In particular, the transport properties[1,2] of nuclear matter, especially the nature and the magnitude of nuclear viscosity, are in the focus of intensive research since many years.

When we disregard the influence of gravity that stabilizes neutron stars, dense objects of nuclear matter can only exist in the form of atomic nuclei with up to about 300 nucleons. Larger nuclei are unstable against binary fission due to the repulsion of the constituent protons. Thus, investigations on the transport coefficients of nuclear matter can only be performed by studying large-scale collective motions of nuclei, whereby fission is an excellent example. Mesoscopic systems, such as atomic nuclei, are small enough that quantum-mechanical features – typical for microscopic systems – are important. However, they are large enough that their behaviour can be understood to a certain extent by using macroscopic concepts. These are a multi-dimensional space of collective degrees of freedom, a thermodynamic behaviour of intrinsic excitations and others. This has decisive consequences for the study of the transport properties of nuclear matter.

The nature and the magnitude of nuclear viscosity are connected with the question, whether nuclei behave like water drops, where collective motions are underdamped, or like honey drops, where collective motions are overdamped. Underdamped systems are mostly governed by inertial forces. They may follow an oscillatory motion around statistical equilibrium for a longer time. The oscillation arises from converting part of their available energy to kinetic energy of the accessible collective degrees of freedom. Overdamped systems are mostly governed by friction. They tend to evolve towards statistical equilibrium of the available states at any moment of a dynamical process with a certain delay, given by the relaxation time[3] .

The dynamical processes that are connected with the fission-fragment mass distributions have been studied in great detail by use of stochastic models, in particular the Fokker-Planck equation[3] and the Langevin approach[4], respectively. The authors also introduced the term of the relaxation time, connecting it with the inverse frequency of the oscillations of an underdamped system and with the characteristic damping time of the motion towards statistical equilibrium of an overdamped system.

**Dissipative and inertial forces in fission**

Dissipation is the process that damps the collective motion of a viscous object. It converts part of the kinetic energy of the object into intrinsic excitations in an irreversible process. Dissipation appears in many scenarios. Fission plays a special role for studying nuclear dissipation for several reasons: (I) Due to the long mean-free path of the nucleons, one-body dissipation is the dominant damping mechanism in low-energy fission[a]. Therefore, fission is ideal for testing the widely used wall- and-window formula[5] for one-body dissipation. (ii) Fission at low excitation energies offers the opportunity for studying dissipation under the influence of pairing correlations[6]. (iii) Fission provides manifold experimental signatures for deducing the influence of dissipation and for determining its magnitude along the entire fission process. Fission cross sections[4], the number of neutrons, light charged particles and gamma rays emitted before scission[7], fluctuations in the fragment properties[4], and odd-even effects in the fragment yields[8] have been exploited.

On the theoretical side, many existing models of nuclear dissipation are essentially classical in nature. There exists a major conflict about the magnitude of the dissipation described by the wall formula[9]. Blocki et al.[5] claim that their formulation of the one-body dissipation is complete and compatible with quantum mechanics. Concerns were raised by Griffin et al.[10] about additional quantum-mechanical effects and by Pal et al.[11] about incomplete chaoticity. Both are assumed to reduce the estimated dissipation considerably. Despite the effort invested experimentally, conclusions remain unclear, often inconsistent, if not contradictory[7].

Fission of heavy nuclei is enabled by the tendency for reducing the energy that is stored in the Coulomb repulsion between the constituent protons. The dynamical evolution of the fissioning system is driven by the potential-energy landscape, which is the potential energy of the fissioning system as function of its shape. Once the fission barrier is passed, the system tends to larger elongations and, finally, to the emergence of two independent fragments. The motion towards scission ends up in an almost infinite number of final states in the two divided fission fragments. Thus, the fissioning nucleus is an open system. However, the fission process is accompanied by many other processes that change its properties, for example the shape and the inner structure, of the fissioning system. All of them are fed by the released Coulomb energy, and, thus, they are orthogonal to the fission direction[12]. The transport coefficients can be very different depending on the shape of the fissioning nucleus (see Appendix). In this work, we focus on the evolution of the elongation and in the mass-asymmetric degree of freedom, respectively. The latter is connected with the appearance of pear-a One-body dissipation, which is realized by the transport of single nucleons (described by the window formula) or by reflections of single nucleons at the nuclear surface (described by the wall formula), is believed to prevail at relatively low energies. This situation is typical in the present context. Two-body dissipation, which is realized by nucleon-nucleon collisions, prevails at higher energies.

like shapes, which result in the formation of fragments with different masses.

The measured mass distribution is the result of a complex dynamical process. The time scale is governed by the transport coefficients in the elongation degree of freedom[b]. On the way to scission, the potential with respect to mass-asymmetric distortion evolves, and the corresponding transport coefficients determine, in which way the asymmetric distortions react to the variations of the potential. This reaction is treated in very different ways by different models. This reaches from an immediate realization of statistical equilibrium[13], corresponding to a relaxation time that is much smaller than the saddle-to-scission time, to a relaxation time that is comparable or larger than the dynamical saddle to scission time[3].

**Abnormal fission yields**

In the course of an intense search for relevant experimental information on the transport properties of nuclear matter, we evidenced abnormal effects in the fragment yields from the fission of light actinides. An anomaly was already observed in the 1970's in the fission excitation functions of several thorium isotopes[14,15,16,17,18,19,20] and attributed to the presence of a third minimum in the fission barrier[21,22,23,24,25,26]. The abnormal behaviour in the fragment yields has not attracted attention before. In a dedicated publication, we pointed the impact of these anomalies on nuclear technology[27]. In this work, we stress the importance for deducing crucial information on the transport properties of nuclear matter.

Figure 1 shows that in almost all fissioning systems the fragment-mass distributions from spontaneous fission, thermal-neutron-induced fission and 14-MeV-neutron-induced fission are very similar. For higher neutron energies, the main difference is the higher yield near symmetry. This effect is well known due to the enhanced yield of the symmetric fission channel, which is favoured by the macroscopic potential. The observed similarities indicate that the shell effects, which are behind the appearance of the asymmetric components in the mass distribution, are rather insensitive to a variation of the excitation energy of the fissioning system in almost all cases. The theoretically expected reduction of the shell strength with increasing excitation energy does not have an influence on the shape of the asymmetric components. Only the data for spontaneous fission of $^{238}U$ differ drastically from the data of neutron-induced fission (thermal to 14-MeV neutrons): A possible explanation of these differences, which is corroborated further down in this work, may be a loss of flux of fission events with heavy fragments near the doubly-magic $^{132}Sn$ and the corresponding light fragments. (The over-shoot of the yields from spontaneous fission around A=92 and A=142 result from the normalization of the yield distribution to 200%.) This suggests that there is a particular effect in the dynamics of the fission process in this system appearing at the lowest excitation energy.

Figure 2 reveals that this suppression effect starts to be seen also in the total kinetic energies in neutron-induced fission of $^{238}U$ at the lowest incident-neutron energies, corresponding to an excitation energy around 6 MeV. Additional data points at lower excitation energies would certainly be appreciated to consolidate this effect. On the contrary, there is no suppression seen in the fission of $^{240}Pu$, even from its ground state by spontaneous fission.

b It is assumed that the shape parametrization is chosen in a way that non-diagonal terms of the tensor of the transport coefficients are negligible.

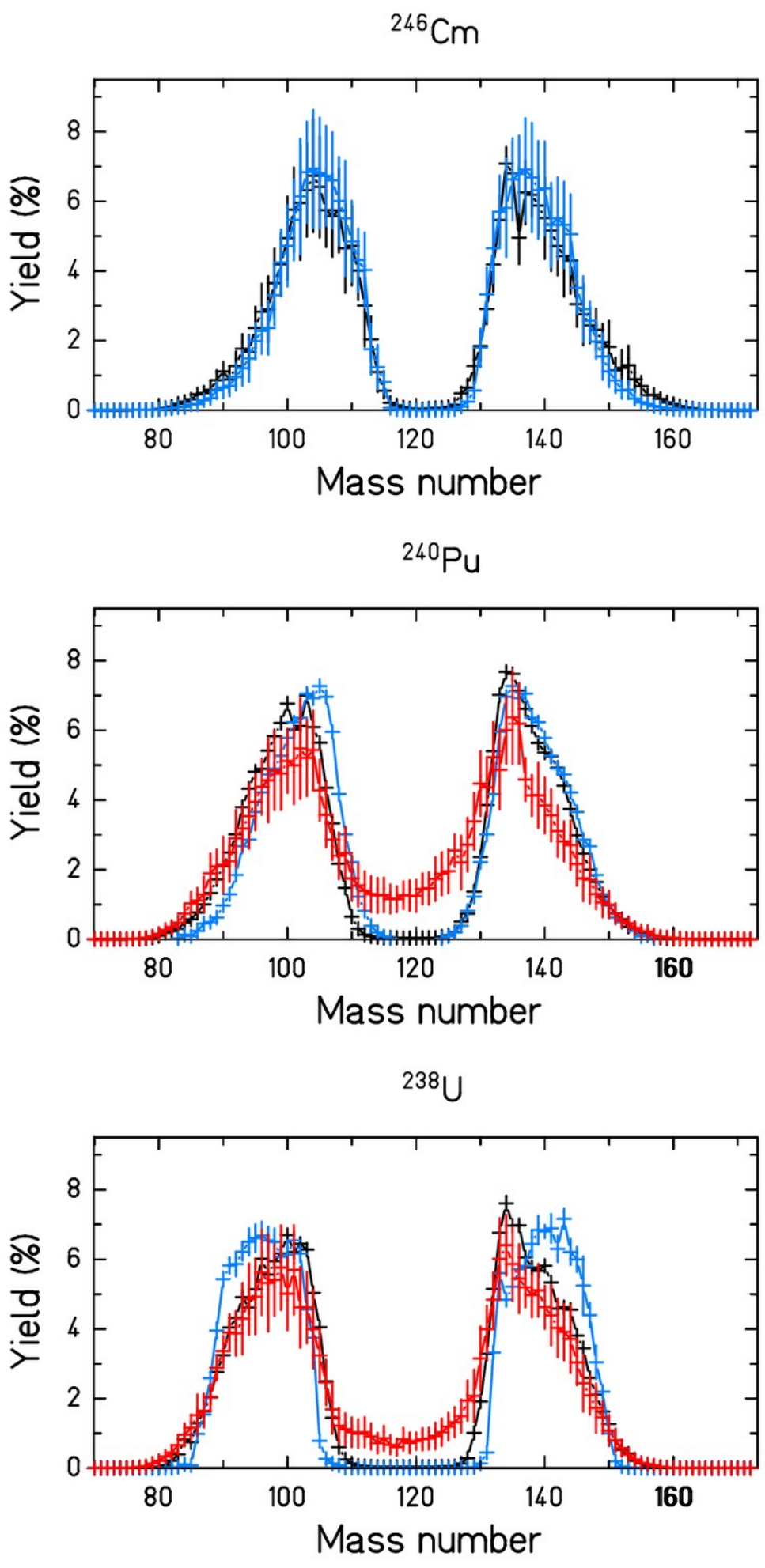

**Fig. 1: Illustration of the anomalous shape of the asymmetric peaks in the mass distribution from spontaneous fission of $^{238}$U.** The inner parts of the asymmetric peaks are strongly suppressed, in comparison with the data from the fission of $^{246}$Cm and $^{240}$Pu. (Data from ENDF/B-VII[28] and [29])
Blue symbols: spontaneous fission, black symbols: fission induced by neutrons below 2 MeV, red symbols: fission induced by 14-MeV neutrons. (In some cases, the data from nearby nuclei are shown, due to the lack of available data, see Table 1.)
All data are post-neutron yields. Only the data from $^{240}$Pu(sf) are pre-neutron yields. This explains the slight shift to higher masses.

**Table 1: List of the fissioning systems used in figure 1.**

| Nominal system | $^{238}$U | $^{240}$Pu | $^{246}$Cm |
|---|---|---|---|
| Spontaneous fission | $^{238}$U(sf) | $^{240}$Pu(sf) | $^{246}$Cm(sf) |
| Low-energy neutrons | $^{238}$U($n_{fast}$,f), $E_n \approx 2$ MeV | $^{239}$Pu($n_{th}$,f),<br>$E_n$ = thermal | $^{245}$Cm($n_{th}$,f),<br>$E_n$ = thermal |
| High-energy neutrons | $^{238}$U(n,f), $E_n$ = 14 MeV | $^{239}$Pu(n,f), $E_n$ = 14 MeV | No data available |

Due to the lack of data, in some cases, the values of the nominal systems are not available. Data from nearby systems are shown in these cases. This choice does not have any sizable influence on the conclusions of the present work.

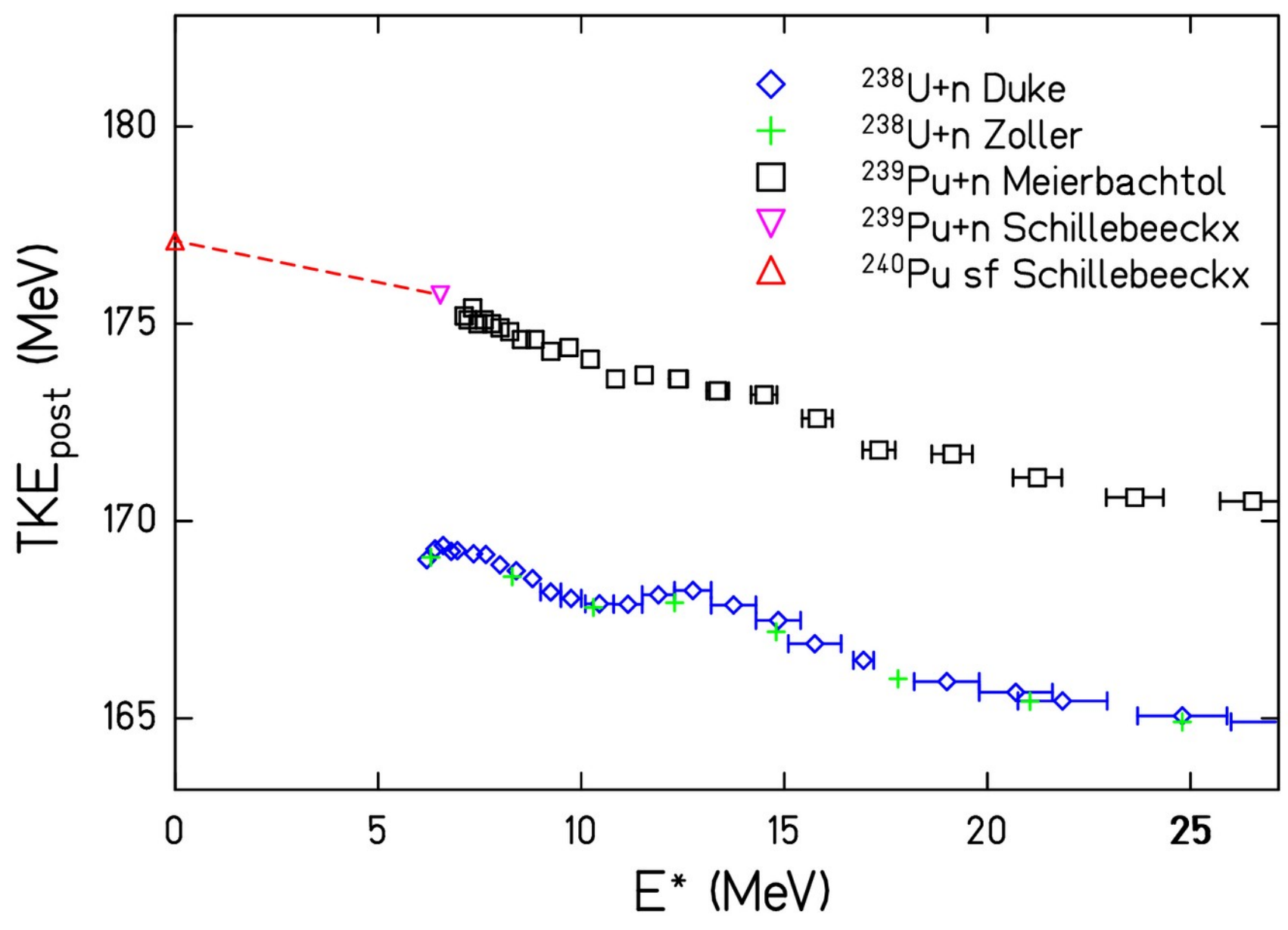

**Fig. 2: Evidence for the suppression effect in TKE.** Measured mean total kinetic energies (TKE) after prompt-neutron emission of the fragments from fission of the compound nuclei $^{239}$U and $^{240}$Pu from refs. [30], [31], [32], [29], (neutron-induced and spontaneous fission). The values for spontaneous fission of $^{240}$Pu and thermal-neutron-induced fission of $^{239}$Pu have been estimated by normalizing the pre-neutron TKE from ref. [29] to the extrapolation of ref. [32] at $E_n$=0.586MeV down to thermal energy. The dashed line is drawn to guide the eye. While the TKE from the fission of $^{240}$Pu shows a continuous increase with decreasing excitation energy down to spontaneous fission, the TKE from the fission of $^{239}$U shows a bending down at the lowest excitation energies. This bending is consistent with the suppression of events with heavy fragments near the doubly magic $^{132}$Sn in spontaneous fission of $^{238}$U seen in Fig. 1, because these events are known to have the highest TKE values.

A similar dependence of the shape of the fragment-yield distribution on the excitation energy was pointed for fission of $^{233}$Pa, induced by protons on $^{232}$Th, see Fig. 3. There, regular oscillations of the peak positions in the mass distributions appear. The peaks of the asymmetric components move towards larger asymmetry, when a sizable part of the fission events is expected to occur from lighter Pa isotopes at energies close to the fission barrier after the pre-fission emission of one or several neutrons. This brings up the question about the origin of the suppression of the fission yields in the inner wings of the asymmetric peaks in these specific cases. In our opinion, the answer is provided by Fig. 4, which shows the evolution of the potential energy along the fission path for a series of systems

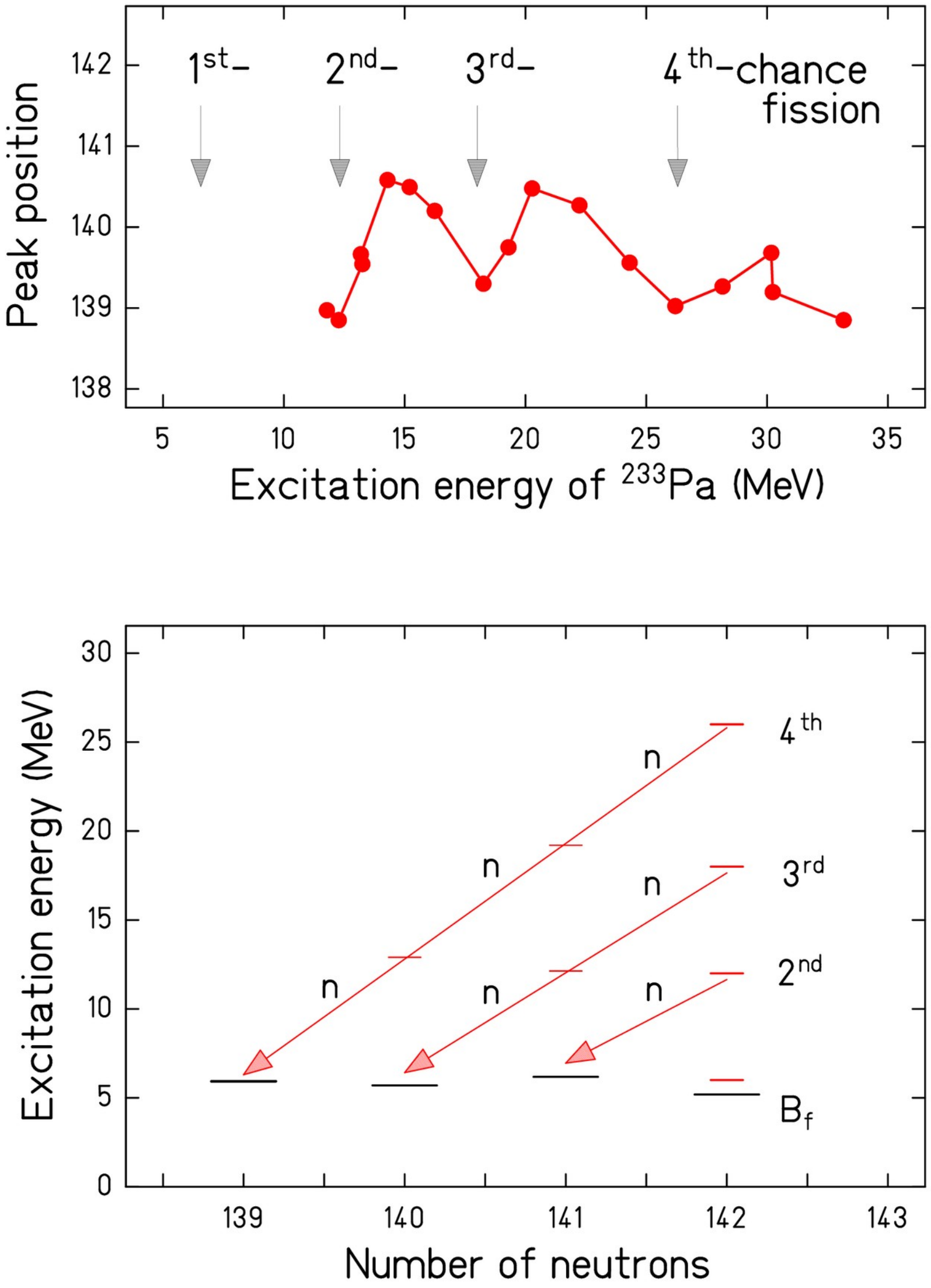

**Fig. 3: Manifestation of the suppression effect in multi-chance fission of $^{233}$Pa**. The u*pper panel* shows the position of the heavy component of the fission-fragment mass distribution from the reaction p + $^{232}$Th as a function of the excitation energy *E** of the compound nucleus $^{233}$Pa, measured by Berriman et al. [33]. The peak position oscillates and shows minima at *E** = 12.5, 18 and 26 MeV. The *lower panel* shows the maximum energies (by short horizontal red lines) of the states in $^{232}$Pa (*N* = 141), $^{231}$Pa (*N* = 140) and $^{230}$Pa (*N* = 139), respectively, that are populated by the emission of one to three neutrons, starting from initial states at the above mentioned energies in $^{233}$Pa (also denoted by short horizontal red lines). They are obtained by subtracting the respective neutron separation energies. Apparently, the final states in $^{232}$Pa, $^{231}$Pa and $^{230}$Pa after the emission of one, two and three neutrons, respectively, are close to the heights of the respective fission barriers. Thus, the minima observed in the upper panel correspond to the thresholds of 2$^{nd}$, 3$^{rd}$ and 4$^{th}$ –chance fission, which are also depicted in the upper panel. The increase of the mean fragment masses above the kink is consistent with the appearance of fission events from low excitation energies at the onset of the next-chance fission with the suppression of compact shapes by the third barrier that we postulate in the present work.

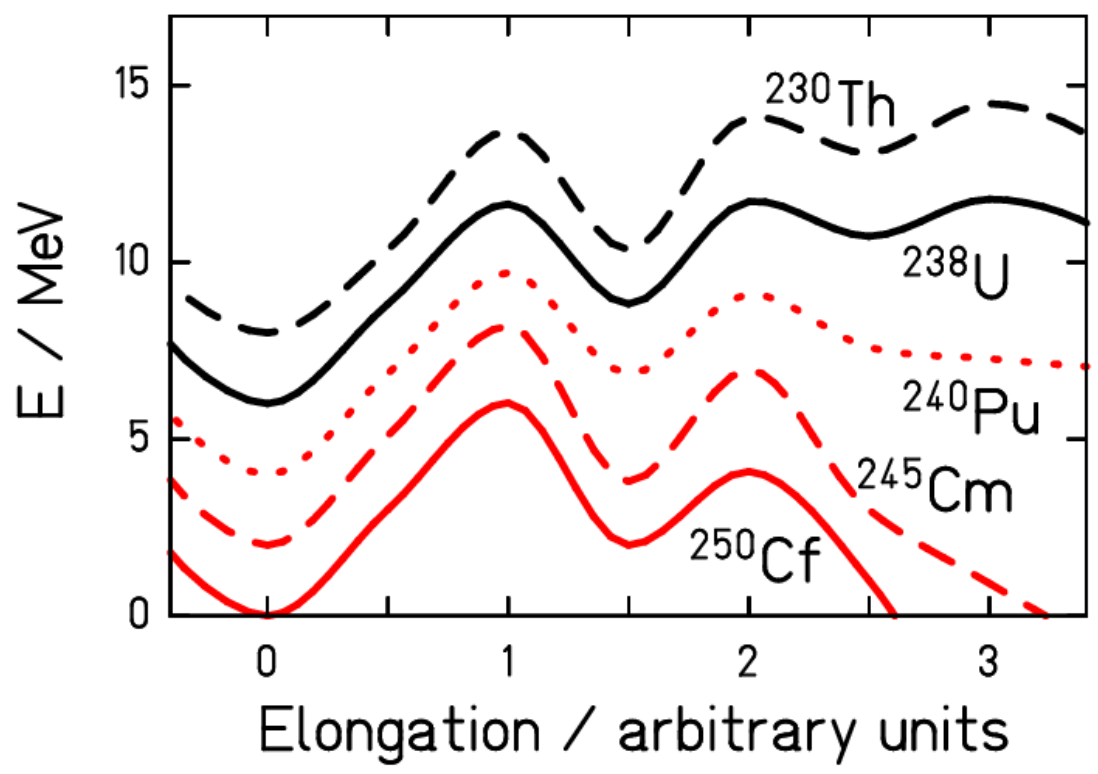

**Fig. 4: Schematic drawing of the potential energy along the fission path.** The curves for *Z*<98 are displaced by ΔE=(98-*Z*)MeV for better visibility. The drawing is based on experimental information about the heights of the first and the second barrier above the ground state (elongation = 0) as well as on the depth of the second minimum, see Table 2. The oscillations that form the multi-humped fission barrier are very similar, while the large-scale (or macroscopic) behaviour that is described by the nuclear liquid-drop model shows a trend from a broad barrier in $^{230}$Th to an appreciably narrower one in $^{250}$Cf. This trend leads to the appearance of a third barrier with comparable height to the first and the second barrier in the lighter systems.

**Table 2: Potential energy along the fission path.**

| System ▶ | $^{230}$Th ▼ | $^{238}$U ▼ | $^{240}$Pu ▼ | $^{245}$Cm ▼ | $^{250}$Cf ▼ | Reference ▼ |
|---|---|---|---|---|---|---|
| Configuration ▼ | | | | | | |
| Ground state | 0 MeV | 0 MeV | 0 MeV | 0 MeV | 0 MeV | Reference energy |
| First barrier | 5.7 MeV | 5.64 MeV | 5.7 MeV | 6.19 MeV | 6.02 | GEF |
| Second minimum | (3.0 MeV) | 2.5 MeV | 2.5 MeV | 1.8 MeV | (1.4 MeV) | Van. 1977[34] |
| Second barrier | 6.09 MeV | 5.71 MeV | 5.09 MeV | 4.95 MeV | 4.08 MeV | GEF |
| Third barrier | (6.48 MeV) | (5.78 MeV) | (3.27 MeV) | (-1.08 MeV) | (-8.66 MeV) | |

The values refer to specific configurations. They are used in Fig. 4. Values in parentheses are estimated by interpolation or extrapolation of empirical trends. Values from GEF[35] are based on the topographic theorem with macroscopic barriers from Thomas-Fermi calculations and empirical ground-state shell effects. They have been validated by measured fission probabilities.

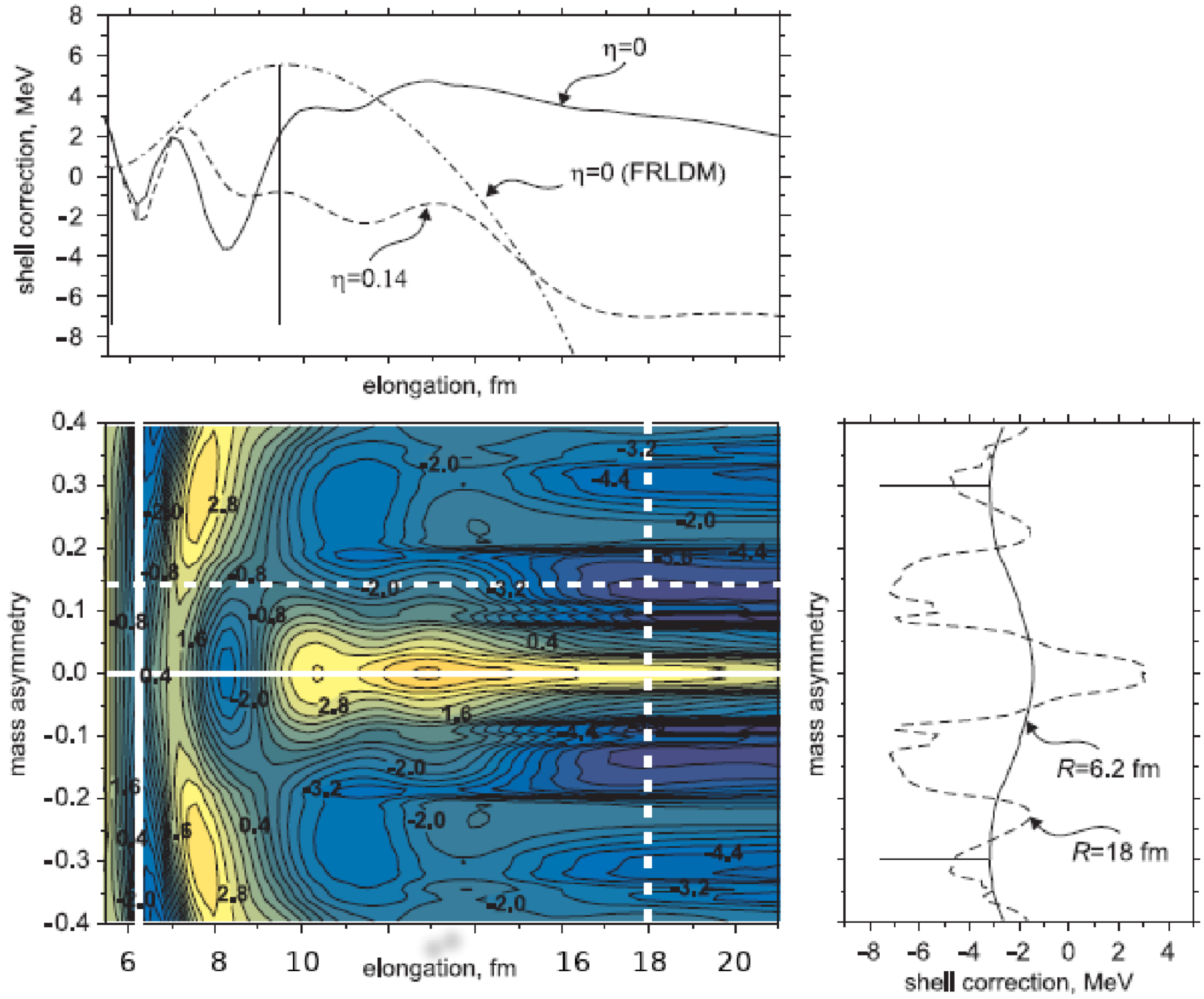

**Fig. 5: Shell-correction energy for the $^{238}$U nucleus in the coordinates (*R*,*η*).** The numbers at the contour lines give the values of the shell correction in MeV. The cuts of the contour plot are made at constant values of the elongation *R* (6.2 fm and 18 fm) and the mass asymmetry *η* (0 and 0.14) shown by thick white lines. Types of the lines correspond to those of the cuts. The three humps of the dashed line in the upper graph appear at the first (*R* = 7 fm), second (*R* = 9.4 fm) and third barrier (*R* = 13 fm), respectively, for the mass asymmetry *η* = 0.14. The dash-dotted curve on the upper graph is the macroscopic potential for symmetric mass split *η* = 0. (The figure is adapted from ref. [37].)

The observed anomaly appears when the system encounters a third barrier with an excitation energy that falls below the third barrier or close to it. The suppressed region in the inner wings of the asymmetric peaks corresponds to very compact configurations with a spherical heavy fragment around the doubly-magic $^{132}$Sn, in agreement with the total kinetic energy signature, see Fig. 2. It is known that for mass splits in this region the total kinetic energy of the fragments, which is strongly fed by the Coulomb energy in the scission configuration, exploits the Q value almost completely[36]. Therefore, it is expected that the potential at the third barrier is already exceptionally high in this mass range. This makes it rather plausible that the yields are very sensitive to an eventual variation of the shape of the potential in direction of mass-asymmetric distortions between the second and the third barrier.

Figure 5 shows the calculated shell effects of the nucleus $^{238}$U from a macroscopic-microscopic model[37] on the two-dimensional plane of elongation and mass asymmetric distortions. The tendency for a slight shift of the negative shell correction towards increased mass asymmetry is clearly seen at the elongation of the third barrier ($R$ = 13 fm). This small shift can have substantial impact on the final yields. The anomalies in the fission yields depicted in Fig. 1, 2 and 3 have a plausible explanation in the suppression of fission events with trajectories passing through compact shapes near scission. We propose the following scenario, which is based on an intuitive understanding of the fission dynamics in the spirit of the multidimensional Langevin dynamics, a well established approach in the field. We refer to the Appendix for a deeper discussion. The suppression effect for splits with heavy fragments around $^{123}$Sn appears, because the system runs out of energy due to the high potential energy at the third barrier for the compact configurations that are typical for these mass splits. The system can only proceed to scission by tunneling, and this reduces the probability for populating fragments around $^{132}$Sn drastically. At higher initial excitation energy, the system can cross the third barrier without tunneling, even in compact configurations. This behaviour provides information on the relaxation time of the mass-asymmetry degree of freedom in comparison with the saddle-to-scission time. If the relaxation time is short, the system will adapt to the potential at the third barrier, and we see the same narrow mass distribution with the suppression of heavy fragments around $^{132}$Sn also at higher excitation energies. If the relaxation time is long, the system will retain the mass-asymmetry distortions acquired at or slightly after passing the second barrier.

**Transport properties of nuclear matter**

Based on the multiplicity of pre-scission neutrons, light charged particles and gamma rays, Hilscher and Rossner (Ref. 7, page 544) deduced a saddle-to-scission time in low-energy fission in the order of $10^{-19}$ to $10^{-20}$ seconds. Although the values determined by other methods do not agree in detail, most of them fall into this rather large time span. Hilscher and Rossner also conclude that this time span indicates that the motion from saddle to scission is overdamped. But, as noted above, this does not mean that the motion in all other degrees of freedom is overdamped, too.

The inertial mass for mass-asymmetric distortions varies considerably on the way from saddle to scission [38] [39], [40], [41]. Even in a macroscopic approach, it tends to diverge close to scission, when the neck diameter shrinks (see Ref. [42], page 29), due to the large velocities of the few nucleons required to establish a certain flux across the neck. This means that the motion along the mass-asymmetry degree of freedom is considerably slowed down with increasing elongation. According to the yield anomalies discussed above, this leads us to conclude that the relaxation time of asymmetric shape distortions between the outer barrier and scission is appreciably longer than $10^{-20}$ seconds. An analogy can be made with the explanation for the width of the fission-fragment distribution in the *N/Z* degree of freedom by a freeze-out of the zero-point motion in the giant isovector dipole resonance before scission due to the strongly increasing inertial mass in this degree of freedom[43]. As a consequence of the increasing inertia, the frequency of the resonance decreases so much that the width of the *N/Z* distribution cannot adjust any more before scission to the adiabatic value that corresponds to the local statistical equilibrium.

The memory of the system on the potential energy at the second barrier revealed in our work may be caused by the same mechanism in mass-asymmetric distortions. We note that theoretical estimations for the nucleus $^{224}$Th [44] show a sharp increase of both the inertia and the friction tensor at large mass-asymmetry at elongations slightly beyond the macroscopic fission barrier. Thus, also some influence of the increasing friction may be expected.

We would like to remind the reader that, in our interpretation, the anomalies in the fission yields found in the fission of light actinides provide a novel and clear evidence for the magnitude of the relaxation time in the mass-asymmetry degree of freedom in comparison with the dynamical saddle to scission time. As discussed in Ref. [45], long relaxation times may occur in an under-damped system due to large inertia or in an over-damped system due to large dissipation. For distinguishing these two scenarios, additional experimental information is required. Nevertheless, the consequences of our finding for these two cases can be stated.

In case of the underdamped scenario, the theoretical models that disregard the influence of inertial mass on the fission dynamics, for example by using the Smoluchowsky equation[46] or by assuming statistical equilibrium at scission, would be in question. Although they can reproduce the main features of the measured fission yields, a detailed description still remains a challenge to them. The anomaly discussed in this work has been overlooked so far, because it hardly exceeds the uncertainties of these calculations.

In case of the overdamped scenario, the anomalies show that the saddle-to-scission time is too short for establishing equilibrium in the mass-asymmetry degree of freedom. Thus, statistical scission-point models would not be justified.

A comprehensive quantitative dynamical calculation of fission at low excitation energy, especially when part of the trajectory consists of a tunneling process, challenges current theory. No consistent calculation exists. As a stopgap, Sadhukhan and collaborators[47] used a hybrid model for calculating fragment yields. They combined the transmission through a barrier by tunneling with classical Langevin dynamics beyond the saddle. Such kind of calculation could help to deduce quantitative conclusions on the transport coefficients of nuclear matter. However, such an undertaking is beyond the scope of this work.

**Conclusion**

Our work provides proof for long relaxation times in the mass-asymmetry degree of freedom, based on the sign of memory in the evolution of mass asymmetry in the fission dynamics between the second barrier and scission. This finding is in line with previous investigations, where a perceptible influence of memory has been deduced from a comparison of measured fragment-mass distributions with the result of transport calculations[4] on the basis of the Langevin equations. However, the interpretation presented in Ref. 4 is less clear compared to our work, because it is based on the excess of the width parameter with respect to a model calculation, while we refer to a new and very specific experimental signature.

In a general sense, the improved knowledge of the transport coefficients established by our analysis provides new constraints on microscopic models that aim to describe the transport properties of nuclear matter, in particular in large-scale collective motion.

**Author information** Correspondence and requests for materials should be addressed to K.-H. S.
(schmidt-erzhausen@t-online.de).

**Acknowledgements** We thank Ewald Gold for carefully reading the manuscript. The work was supported by the French-German collaboration between IN2P3-DSM/CEA and GSI, under Agreement No. 19-80. AH is grateful for funding from the Knut and Alice Wallenberg Foundation under KAW 2020.0076.

# Appendix

## Stochastic differential equations

The terms „potential energy”, „inertial mass“ and „viscosity“ are well known for describing the dynamical processes of massive objects in our environment. The corresponding potential, inertial and friction forces form the decisive ingredients of the Fokker-Plank differential equation[48] or the equivalent Langevin equations[49], [c]. The inertial mass determines the kinetic energy of an object that moves with a certain velocity. The potential energy is the work that has to be invested for displacing the object against a conservative force. This means that the energy can be retrieved by inverting the displacement. Viscosity is the property of a medium that slows down the velocity of a moving object by dissipating its ordered energy (e.g. the kinetic energy) into a disordered form (e.g. heat) in an irreversible process. Viscosity has an intimate connection with a random force, which is expressed by the Einstein relation. It adds fluctuations to the dynamics.

## Modeling of fission dynamics

The appropriate microscopic approach for the description of a nucleus would be to consider the forces that act between any nucleon with all the others and to solve the appropriate differential equations of motion. This is a tremendously complex task. The problem may be simplified by considering that a nucleus has several properties, which resemble those of a classical drop of a liquid. Some of those are a constant volume, a well defined surface, and a self-consistent adjustment of the shape to the acting forces. In this context, the classical Langevin equations are widely used, and particularly successful, for describing large-scale collective nuclear motions, including fission, although this is not obvious. We here illuminate the abstractions, assumptions and simplifications, which must be applied for employing such an approach that has been developed for the motion of a point-like object in a viscous medium[49] . The introduction of this relatively simple approach favours an intuitive understanding of the problem tackled in the present work.

## Multi-dimensional deformation space

In this approach, the potential energy that depends on the shape of the nucleus is commonly expressed by introducing an abstract multidimensional deformation space. The shape is defined by distortions of a sphere, for example by extending or compressing it in one direction to an ellipsoid. Triaxial shapes are introduced by considering an ellipsoid with three different main axes. Also a pear-like shape can be chosen as additional distortion. In principle, an infinite number of distortions can be defined. A specific, more or less complex, shape corresponds to a point in the multidimensional deformation space. The potential energy at a specific point may be calculated by the microscopic approach mentioned above. For the motion of the point in deformation space, representing a change of the shape, the displacement of at least some nucleons of the nucleus is required. This is connected with a certain kinetic energy. This defines the inertial mass, associated with a certain kind of distortion. However, this definition is not unique. Any modification of the positions of the nucleons that end up in the same shape is possible. Here an additional condition, for example the motion of the nucleons with the lowest associated kinetic energy (see ref.[50] about “irrotational” and “rotational flow”), makes the definition of the inertial mass unique. If the motion of the nucleons is accompanied by friction, dissipation arises. The magnitude of inertia and dissipation is given by their respective transport coefficient. The inertial mass and the dissipation coefficients may depend on the specific kind of shape distortion, the value

c It is not our intention here to introduce the Fokker-Planck equation or the Langevin equations in full detail. For this purpose, we refer to the dedicated literature. Our aim is to introduce the abstract deformation space and the most relevant forces that determine nuclear collective dynamics.

of the corresponding distortion as well as on temperature and other parameters. This makes the dynamics of the system in the multi-dimensional deformation space rather complicated.

**Application to nuclear fission**

A fissioning nucleus is an open system. That means, it is not bound with respect to the elongation degree of freedom. Instead, the motion towards scission ends up in an almost infinite number of final states in the two fission fragments. However, the fissioning nucleus is bound in any other direction in the multi-dimensional deformation space. In most directions, this can be represented by a parabola-like potential, modulated by quantum-mechanical shell effects[d]. Strong shell effects favour mass-asymmetric distortions on the way to scission that lead to fission fragments close to specific numbers of protons and/or neutrons. This is well illustrated by the double-humped mass distributions, shown in Fig. 1 of the main text.

The present work deals with the motion in mass-asymmetric distortions during the directed motion towards scission. The crucial question is, whether the distribution in mass-asymmetric distortion adapts quickly to the shape of the corresponding potential in such a way that the population of states in mass asymmetry matches local equilibrium at any elongation. Our analysis shows that the distribution in mass asymmetry at or close to the second barrier persists up to scission, as long as the influence of tunneling beyond the second barrier is negligible.

---

d Quantum-mechanical shell effects mean a grouping of the energies of stationary states in a potential pocket. There are similarities to standing waves with distinct frequencies of an elastic rope that is fixed on both ends.

1 Nadtochy, P. N., Ryabov, E. G., Karpov, A. V. & Adeev, G. D. Transport coefficients for modeling nuclear fission dynamics. *Comp. Phys. Comm.* **275**, 108308 (2022).
2 Okada, K., Wada, T. & Carjan. N. Influence of transport oefficients on fission dynamics in the five-dimensional Langevin approach. *Phys. Rev. C* **112**, 034613 (2025).
3 Adeev, G. D. & Gonchar, I. I., The Dynamical Description of the Mass Distribution of Fission Fragments. *Z. Phys. A* **320**, 451 (1985).
4 Mazurek, K., Nadtochy, P. N., Ryabov & E. G., Adeev, G. D., Fission fragment distributions within dynamical approach. *Eur. Phys. J. A* **53**, 79 (2017) / Erratum: *Eur. Phys. J. A* **53**, 144 (2017).
5 Blocki, J., Borneh, Y., Nix, J. R., Randrup, J., Robel, M., Sierk, A. J. & Swiatecki, W. J. One-body dissipation and super-viscosity of nuclei. *Annals of Physics* **113**, 330 (1978).
6 Jurado, B. & Schmidt, K.-H. Influence of complete energy sorting on the characteristics of the odd-even effect in fission-fragment element distributions. *J. Phys. G: Nucl. Part. Phys.* **42**, 055101 (2015).
7 Hilscher, D. & Rossner, H. Dynamics of nuclear fission. *Ann. Phys. Fr.* **17**, 471 (1992).
8 Ramos, D. et al. Experimental evidence of the effect of nuclear shells on fission dissipation and time. *Phys. Rev. C* **107**, L021601 (2023).
9 Bender M. et al. Future of nuclear fission theory. *J. Phys. G: Nucl. Part. Phys.* **47**, 113002 (2020).
10 Griffin, J. J. & Dworzecka, M. Quantal one-body dissipation: limitations of the classical wall formula. *Phys. Lett. B* **156**, 139 (1985).
11 Pal, S. & Mukhopadhyay, T. Shape dependence of single particle response and the one body limit of damping of multipole vibrations of a cavity. *Phys. Rev. C* **54**, 1333 (1996).
12 Nix, J. R. The normal modes of oscillation of a uniformly charged drop about its saddle-point shape *Ann. Phys.* **41**, 52 (1967).
13 Bulgac, A., Jin, S. & Stetcu, I. Nuclear fission dynamics: past, present, needs, and future. *Front. Phys.* **8**, 63 (2020).
14 Back, B. B., Britt, H. C., Garrett, J. D. & Hansen, O. Subbarrier fission resonances in Th isotopes. *Phys. Rev. Lett.* **28**, 1707 (1972).
15 Bhandari, B. S. Three-hump fission barrier in $^{232}$Th. *Phys. Rev. C* **19**, 1820 (1979).
16 Blons, J. A third minimum in the fission barrier. *Nucl. Phys. A* **502**, 121c (1989).
17 Csatlos, M., et al. Resonant tunneling through the triple-humped fission barrier of $^{236}$U. *Phys. Lett. B* **615**, 175 (2005).
18 Csige, L. et al. Transmission resonance spectroscopy in the third minimum of $^{232}$Pa. *Phys. Rev. C* **85**, 054306 (2012).
19 Csige, L. et al. Exploring the multihumped fission barrier of $^{238}$U via sub-barrier photofission. *Phys. Rev. C* **87** (2013) 044321.
20 Oberstedt, A. & Oberstedt, S. (2023). The Multi-Humped Fission Barrier. In: Tanihata, I., Toki, H., Kajino, T. (eds) Handbook of Nuclear Physics . Springer, Singapore. https://doi.org/10.1007/978-981-15-8818-1_79-2.
21 Cwiok, S., Nazarewicz, W., Saladin, J. X., Plociennik, W. & Johnson, A. Hyperdeformation and clustering in the actinide nuclei. *Phys. Lett. B* **322**, 304 (1994).
22 Jachimowicz, P., Kowal, M. & Skalski, J. Eight-dimensional calculations of the third barrier in $^{232}$Th. *Phys. Rev. C* **87**, 044308 (2013).
23 Ichikawa, T., Möller, P. & Sierk, A. J., Character and prevalence of third minima in actinide fission barriers. *Phys. Rev. C* **87**, 054326 (2013).
24 McDonnell, Nazarewicz, W. & Sheikh, J. Third minima in thorium and uranium isotopes in a self-consistent theory. *Phys. Rev. C* **87**, 054327 (2013).
25 Schmitt, C., Pomorski, K., Nerlo-Pomorska & B., Bartel, J. Performance of the Fourier shape parametrization for the fission process. *Phys. Rev. C* **95**, 034612 (2017).
26 Okada, K., Wada, T., Capote, R. & Carjan, N. Cassini-oval description of the multidimensional potential energy surface for 236U: Role of octupole deformation and calculation of the most probable fission path. *Phys. Rev. C* **107**, 034608 (2023).
27 Schmidt, K.-H., Schmitt, Ch., Heinz, A. & Jurado, B. Identifying and overcoming deficiencies of nuclear data on the fission of light actinides by use of the GEF code. *Ann. Nucl. Energy* **208**, 110784 (2024).
28 Chadwick, M. B., ENDF/B-VII.1 nuclear data for science and technology: cross sections, covariances, fission product yields and decay data. *Nucl. Data Sheets* **112**, 2287 (2011).
29 Schillebeeckx, P., Wagemans, C., Deruytter, A. J. & Barthelemy, R. Comparative study of the fragments' mass and energy characteristics in the spontaneous fission of $^{238}$Pu, $^{24}$0Pu and $^{242}$Pu and in the thermal-neutron-induced fission of $^{239}$Pu, *Nucl. Phys. A* **545**, 623 (1992).
30 Duke, D. L., Tovesson, F., Laptev, A. B., Mosby, S., Hambsch, F.-J., Brys, T. & Vidali, M. Fission-fragment properties in U238(n,f) between 1 and 30 MeV. *Phys. Rev.* C **94**, 054604 (2016).

31 P. Zöller, PhD thesis, Techn. Hochschule Darmstadt, Germany (1995).
32 Meierbachtol, K. et al. Total kinetic energy release in Pu239(n,f) post-neutron emission from 0.5 to 50 MeV incident neutron energy. *Phys. Rev.* C **94**, 034611 (2016).
33 Berriman, A., et al. Energy dependence of p + $^{232}$Th fission mass distributions: Mass-asymmetric standard I and standard II modes, and multichance fission. *Phys. Rev. C* **105**, 064614 (2022).
34 Vandenbosch, R. Spontaneously fissioning isomers. *Ann. Rev. Nucl. Sc.* **27**, 1 (1977).
35 Schmidt, K.-H., Jurado, B., Amouroux C. & Schmitt, C., General description of fission observables: GEF model code. *Nucl. Data Sheets* **131**, 107 (2016).
36 Gönnenwein, F. & Börsig, B., Tip model of cold fission. *Nucl. Phys. A* **530**, 27 (1991).
37 Karpov, A. V., Kelic, A. & Schmidt, K.-H. On the topographical properties of fission barriers. *J. Phys. G: Nucl. Part. Phys.* **35**, 035104 (2008).
38 Ivanyuk, F. A., Hofmann, H., Pashkevich, V. V. & Yamaiji, S., Transport coefficients for shape degrees in terms of Cassini ovaloids. *Phys. Rev. C* **55**, 1730 (1997).
39 Ayik, S., Suraud, E., Stryjewsi, J. & Belkacem, M., From one-body to collective transport models. *Z. Phys. A* **337**, 413 (1990).
40 Usang, M. D., Ivanyuk, F. A., Ishizuka, C. & Chiba, S., Effects of microscopic transport coefficients on fission observables calculated by the Langevin equation. *Phys. Rev. C* **94**, 044602 (2016).
41 Usang, M. D.  Ivanyuk, F. A., Ishizuka C. & Chiba, S., Analysis of the total kinetic energy of fission fragments with the Langevin equation. *Phys. Rev. C* **96**, 064617 (2017).
42 Nix, J. R. & Swiatecki, W. J. Studies in the liquid-drop theory of nuclear fission. *Nucl. Phys.* 71, 1 (1965).
43 Nifenecker, H. A. dynamical treatment of isobaric widths in fission : An example of frozen quantal fluctuations. *J. Physique Lett.* **41**, 47 (1980).
44 Adeev, G. D., Karpov, A. V., Nadtochii, P. N. & Vanin, D. V. Multidimensional stochastic approach to the fission dynamics of excited nuclei. *Phys. Part. Nucl.* 36, **378** (2005).
45 Adeev, G. D., Gonchar, I. I., Pashkevich, V. V., Pischasov, N. I. & Serdyuk, O. I., Diffusion Model of the Formation of Fission-Fragment Distributions. *Sov. J. Part. Nucl.* **19**, 529 (1988).
46 Chandrasekhar, S. Stochastic problems in physics and astronomy. *Rev. Mod. Phys.* **15**(1), 1 (1943).
47 Sadhukhan, J., Nazarewicz, W. & Schunk, N. Microscopic modeling of mass and charge distributions in the spontaneous fission of $^{240}$Pu. *Phys. Rev. C* **93**, 011304(R) (2016).
48 Abe, Y., Reinhard, P.-G. & Suraud, E. On stochastic approaches of nuclear dynamics. *Phys. Rep.* **275**, 49 (1996).
49 Langevin, P. Sur la théorie du mouvement brownien [On the theory of Brownian Motion]. *C. R. Acad. Sci. Paris.* **146**, 530 (1908).
50 Nix, J. R., Swiatecki, W. J. Studies in the liquid-drop theory of nuclear fission. *Nucl. Phys.* **71**, 1 (1965).